\documentclass[11pt,preprintnumbers,aps,amssymb,nofootinbib,amsmath,superscriptaddress,notitlepage]
{revtex4-1}
\usepackage{epsfig,epsf}
\usepackage{comment}
\usepackage{bm} 
\usepackage{indentfirst}
\usepackage{color} 
\usepackage{slashed} 
\usepackage{relsize}	
\usepackage{soul} 
\usepackage{hyperref}
\usepackage{tensor} 
\newcommand{\beq}{\begin{equation}}
\newcommand{\beql}[1]{\begin{equation}\label{#1}}
\newcommand{\eeq}{\end{equation}}
\def\bal#1\gal{\begin{align}#1\end{align}}
\newcommand{\ball}[1]{\bal\label{#1}}
\begin{document}

\title{Angular Power Spectrum and Elliptic Flow from Event Maps in Heavy Ion Collisions}

\author{Hannah Anderson}
\author{Shengquan Tuo}
\affiliation{Department of Physics and Astronomy, Vanderbilt University, Nashville, TN 37235, USA}
\date{\today} 

\date{\today}

\begin{abstract}

Azimuthal and polar angle distributions of particles produced in heavy ion collisions carry important information about the early state of the system and the evolution of the Quark Gluon Plasma. The pixelization code HEALPix was created by the Jet Propulsion Laboratory to analyze the cosmic microwave background. As has been shown using public data from the ALICE experiment, its two-dimensional representation of a sphere containing pixels of equal area has a broader application to heavy ion collisions. The angular power spectrum, an application of HEALPix, is directly related to the particle flow and is calculated through the angles theta and phi on the sphere. The elliptic flow is calculated from simulated AMPT events and publicly available CMS data.  We show that HEALPix can be used to detect the presence of elliptic flow for lead-lead collisions and thus has broader applications in the study of QGP in heavy ion collisions.

\end{abstract}

\maketitle

\section{Introduction}\label{sec:a}

Heavy ions colliding at ultrarelativistic energies, as studied at the BNL RHIC and CERN LHC, can form a state of matter coined the Quark Gluon Plasma (QGP) \cite{STAR:2005gfr,PHENIX:2004vcz,Alver:2008aa,BRAHMS:2004adc, Aamodt:2010pa, ATLAS:2012at,Chatrchyan:2012ta}. QGP is formed in the moments of the collisions and contains quarks and gluons as its degrees of freedom. After its initial creation, QGP expands as a collective system denoting that the initial geometry formed by the initial overlap between colliding particles will depict the distribution of the emitted hadrons after the system cools through a phase transition. This expansion can be described through viscous hydrodynamic analysis through which information on QGP’s viscosity can be obtained. In order to analyze this and other transport properties, azimuthal anisotropies have been studied through the flow coefficient \(v_n\) relating to the final hadronic distribution Fourier decomposition. Specifically, the categorization of elliptic flow through the harmonic coefficient \(v_2\) is central to the indication of QGP as it reflects the initial spatial anisotropies of the overlap region in the transverse plane seen at the very beginning of the collective expansion\cite{Luzum:2009sb, Schenke:2010rr, Heinz:2013th}. 
 
 It has been discovered that QGP contains many commonalities between the “Big Bang” at the start of the evolution of the universe, and has thus been coined “Little Bangs”. What we know about the early universe shortly after the Big Bang has been described through the photons emitted from the last scattering when atoms formed. These photons were discovered in 1964 and are denoted Cosmic Microwave Background (CMB) and appear as electromagnetic radiation~\cite{1965ApJ142419P}. From this study, it is clear there was some initial curvature of the universe which has been studied through the angular power spectrum which relies on the anisotropic emission of the photons. Due to the similarities between the Big and Little Bangs, it was proposed in ref. \cite{Naselsky:2012nw} that the angular power spectrum used to study anisotropies in the CMB could be used to analyze large-multiplicity heavy ion collisions. The study would depict the hadrons from the collisions as the photons created through the last scattering, and could be viewed through the angular power spectrum which utilizes spherical harmonics and thus views the distribution of hadrons as a function of the polar angle $\theta$ and azimuthal angle $\phi$. 
 
 In this study, a look into the angular power spectrum is detailed through analysis of simulated and public heavy ion collision data. The datasets focus on Pb-Pb collisions at 2.76 TeV obtained through the CMS experiment collection. The public data is taken from the CERN open data portal~\cite{CMS2010HIAllPhysics}, and the simulations are conducted through both a Monte Carlo toy model and the more intricate A MultiPhase Transport (AMPT) model~\cite{Lin:2004en}. Similar work has been conducted in studies~\cite{Naselsky:2012nw, Sarwar:2017iax, Llanes-Estrada:2016pso,Machado:2018xvi}, and the methods mentioned below draw many similarities to the methods mentioned in~\cite{Machado:2018xvi} with the notable difference in the data collection. Because CMS data is processed in this study, the edge effects due to detector acceptance cut studied in the above paper is all but erased except near the top of the polar caps. This discrepancy is rectified through the subtraction of the m = 0 modes which in this case does not change the results significantly. We are able to compare the results of \(v_2\) obtained through the analysis of the angular power spectrum with the Q-cumulants method, which is a more traditional analysis of flow. In addition, the effects of jets are briefly studied through the use of a simple toy model. The jets are magnified in order to understand the severity of the change caused in the angular power spectrum and the flow model. This is also interpreted through changes implemented in the initial setup of AMPT which denote the presence or absence of elliptic flow through the inclusion of the string melting model.

\section{HEALPix Methodology and Modeling}\label{sec:b}

As the analytic method introduced above is concerned with viewing the distribution in terms of the polar angle $\theta$ and the azimuthal angle $\phi$, the specific distribution needs to be defined. Hadronic particles produced in heavy ion collisions can be described through their distribution across the detector at the LHC and can be described using either the polar angle $\theta$ or the pseudorapidity and the azimuthal angle $\phi$ between the 3-momentum of the particle and its beam access. This has been defined through a distribution density function \(f(\theta, \phi)\) and can be expressed in terms of spherical harmonics
\ball{a2}
f(\theta, \phi) = \sum_{l=0}^{l_{max}} \sum_{m=-l}^{m=l} a_{lm}Y_{l}^{m}(\theta, \phi) \,
\gal

 with,
 
 \ball{aa2}
Y_{l}^{m}(\theta, \phi) = \sqrt{\frac{2l+ 1}{4\pi}\frac{(l-m)!}{(l+m)!}}P_l^m(cos(\theta))e^{im\phi} \,
\gal

  where $\theta \in [0,\pi]\, \phi \in [0,2\pi)\,$ and $P_l^m(cos(\theta))$ are the Legendre Polynomials. The l-modes are set as the multipole moments, and it is assumed that modes with $l> lmax$ hold insignificant power. The spherical harmonics define the contribution of the spherical shape to the overall shape of the distribution

In order to depict the angular power spectrum with these equations, it was important to use simulated data before applying the CMS open data. This was performed through the use of the AMPT model~\cite{Lin:2004en}. This simulation relies on multiple pieces starting with the Heavy Ion Jet Interaction Generator (HIJING) to generate initial conditions which then transfers to the Zhang’s Parton Cascade (ZPC)  to model partonic scattering, Lund string fragmentation model or quark coalescence for hadronizations, and finally A Relativistic Transport (ART) to model hadronic scattering. This combination of models is able to accurately depict heavy ion collisions, and it is composed of very sensitive initial conditions which can be used to change the heavy ion properties before the collision and setup necessary to produce the desired results. In this section, Pb-Pb ions are portrayed with 5.02 TeV with ~200,000 events. The impact parameter is set to be $b_{max} < 10 fm$ with string melting activated, and the reaction plane orientation for each event is randomized. 
\begin{figure}[ht]
\begin{minipage}[b]{0.45\linewidth}
            \centering
            \includegraphics[width=\textwidth]{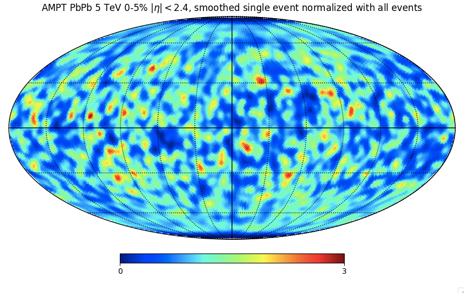}
            \caption{Single event distribution from AMPT normalized by all event distribution. Particles with $p_T > 0.3$ GeV  and $|\eta|<2.4$ are used.}
            \label{fig:a}
        \end{minipage}
        \hspace{0.5cm}
        \begin{minipage}[b]{0.45\linewidth}
            \centering
            \includegraphics[width=\textwidth]{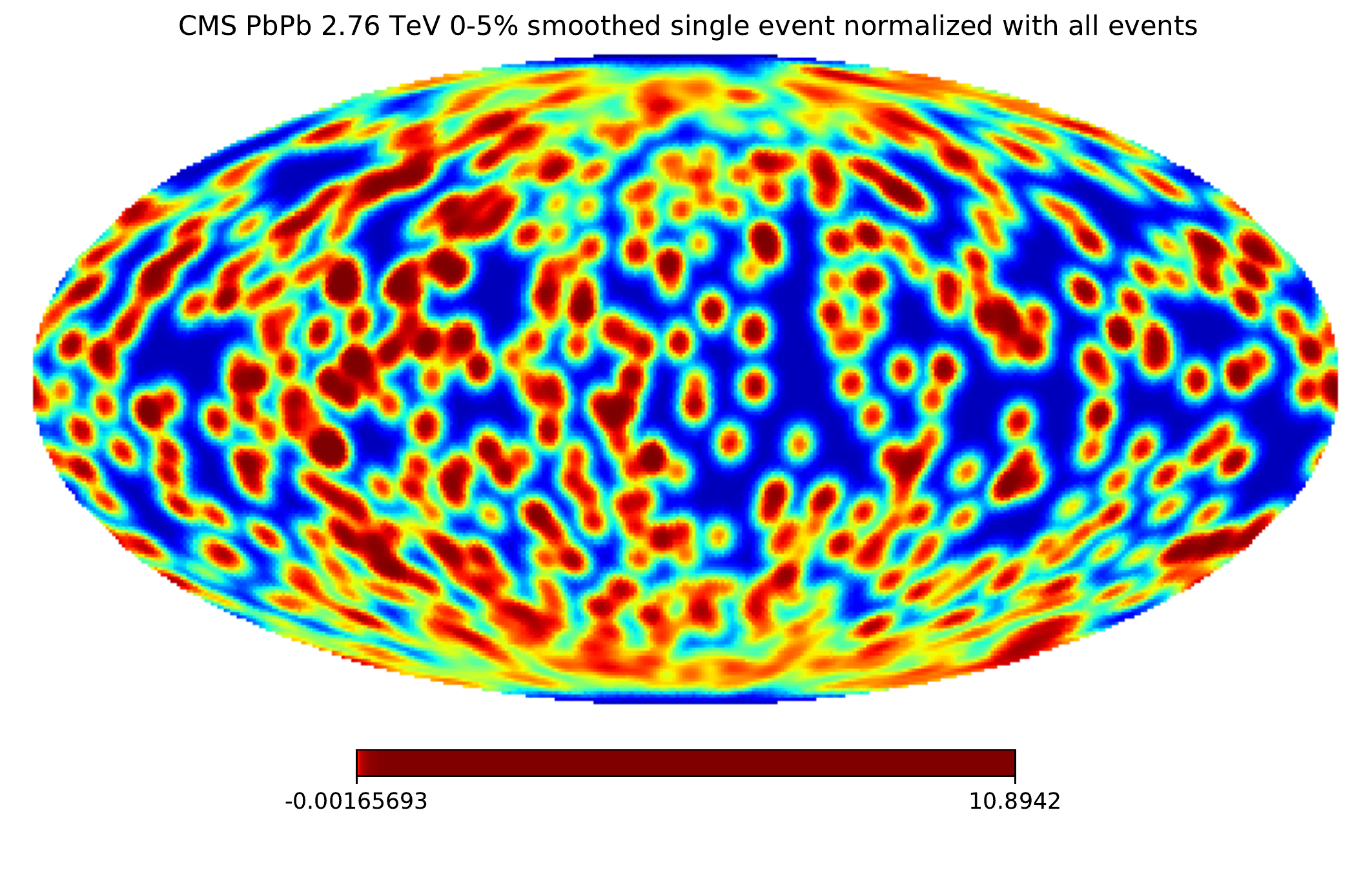}
            \caption{Single event distribution from CMS open data normalized by all event distribution. Particles with $p_T > 0.3$ GeV  and $|\eta|<2.4$ are used.}
            \label{fig:b}
        \end{minipage}
\end{figure}

The software that is used to perform the angular power spectrum analysis is denoted HEALPix (Hierarchical Equal Area isoLatitude Pixelation) \cite{Gorski:2004by} which was created by the Jet Propulsion Laboratory in order to assess the CMB. In order to do so, HEALPix divides a spherical surface into sections of equal area translated into equal area pixels, and are mapped onto a Mollweide map. This map provides a flat distribution of the CMB photonic scattering, or in this case the hadronic particle distribution over the detector. The number of pixels depends on the resolution chosen for the analysis of the angular power spectrum. Through functions described within HEALPix, each position $(\theta, \phi)$ is mapped to a pixel. Each pixel responds then to colors that depict the particle density map times the number of pixels. 

Though there is not much of a cut due to detector limitations in CMS data, the cut is still seen near the top of the polar angles and is defined by a pseudorapidity range. Pseudorapidity relates to the polar angle $\theta$ via $\eta=-ln\big[\tan(\theta/2)\big]$, and the range of $|\eta| < 2.4$ is imposed on the reconstructed particles in both the open data and the simulated AMPT data. The sphere is divided into 768 pixels when performing calculations which corresponds to an $l_{max}$ of 23 when calculating the distribution. This then directly relates to the calculation of the angular power spectrum.

In order to prepare the data to analyze through the angular power spectrum, the detector affects needed to be accounted for as well. This was performed by mapping all events onto a Mollweide map and dividing each successive single event map by the total event map to get rid of anisotropies within the total map that denote detector effects. This map is depicted by $\bar{f}(\theta,\phi) = f(\theta,\phi)/F_{total}(\theta,\phi)$. Figure~\ref{fig:a} and ~\ref{fig:b} shows the $\bar{f}(\theta,\phi)$ map in AMPT and CMS open data with gaussian smoothing applied. 

The calculations of the angular power spectrum is represented by 
\ball{a3}
C_l = \frac{1}{2l+1}\sum_{m=-l}^{m=l}|a_{lm}|^2\,
\gal
and consists of the coefficients $a_{lm}$. $a_{lm}$ coefficients are calculated through the spherical harmonics and the corrected density distribution. 
\ball{aa3}
a_{lm} = \frac{4\pi}{N_{pix}}\sum_{j=0}^{N_{pix}-1}Y_{lm}^*(\theta_j,\phi_j)\bar{f}(\theta_j,\phi_j)\,
\gal
where $N_{pix}$ is the total number of pixels represented in the sphere. 
The representation of the single event is relatively similar to that of the average events with more fluctuations in the higher $l$-modes due to anisotropies that form in each event that can be averaged out. The $C_0$ value is fixed at 4$\pi$ for all the angular power spectrums due to the normalization that is performed in healpix with the corrected event maps.
\begin{figure}[ht]
\begin{minipage}[b]{0.47\linewidth}
            \centering
            \includegraphics[width=\textwidth]{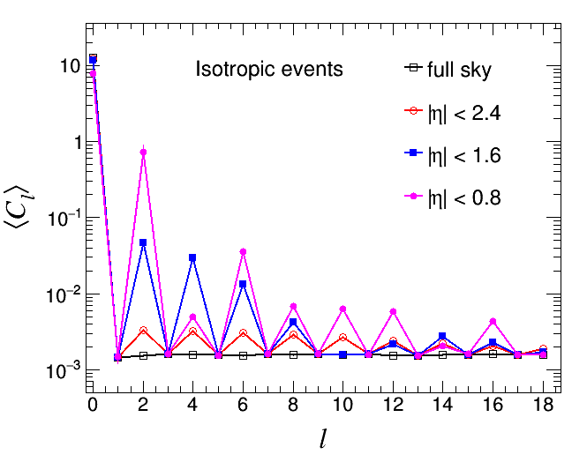}
            \caption{Isotropic data processed over different $\eta$ ranges to demonstrate the edge effects that appear with smaller $|\eta|$ acceptances}
            \label{fig:c}
        \end{minipage}
        \hspace{0.5cm}
        \begin{minipage}[b]{0.47\linewidth}
            \centering
            \includegraphics[width=\textwidth]{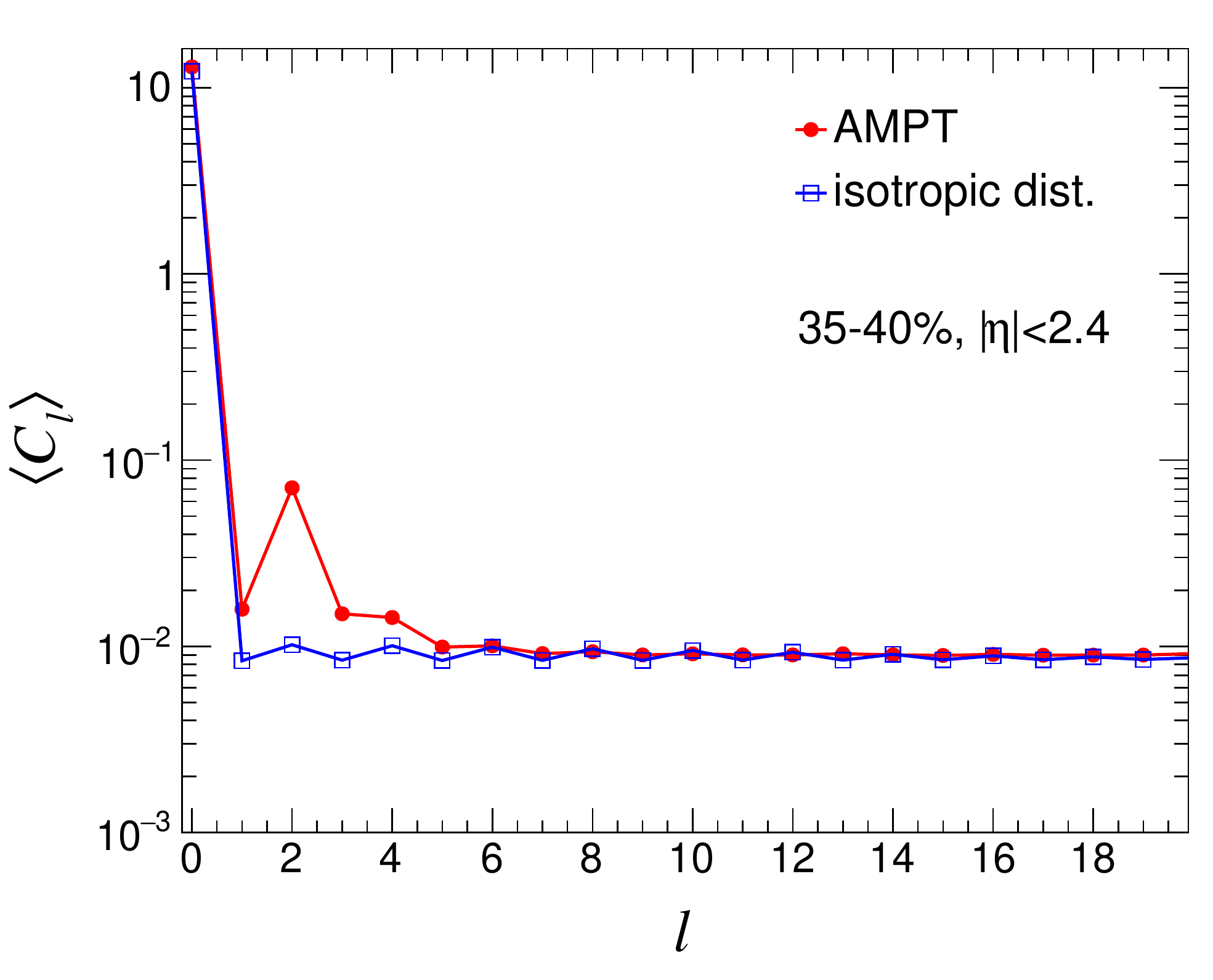}
            \caption{Comparison between AMPT and isotropic distribution. Isotropic distribution has very small contributions of $l$ modes larger than 0}
            \label{fig:d}
        \end{minipage}
\end{figure}

In addition to working with simulated data from AMPT, CERN’s CMS open data was used. The open data only saved particles with $p_T> 0.9$ GeV and $|\eta|<2.4$. The range originally designated in the AMPT model includes $p_T> 0.2$ GeV, but the open data only includes $p_T> 0.9$ GeV which has a larger angular power spectrum with the contribution of higher $l$ modes. This was also compared with AMPT $p_T>0.9$ GeV   which has produced a similar angular power spectrum.

Additionally, there may be edge effects that are applied to the maps and calculations due to the pseudorapidity range that is applied to the data. Though the range covers very nearly the entire sphere, isotropic data will still be analyzed. After mapping ~10,000 events per 5\% centrality bin and averaging the single event angular power spectrums in the 5\% centrality ranges, isotropic data was created with the same multiplicity. When observing the effects that pseudorapidity ranges have on isotropic distributions in Fig.~\ref{fig:c}, it is clear that in the “full sky” relies only on the monopole distribution while the other spherical harmonics in the larger $l$ modes do not play any role. In this graph there are also clear edge effects that occur in smaller pseudorapidity ranges as were described in Ref.~\cite{Machado:2018xvi}.  In Fig.~\ref{fig:d} the AMPT simulation is compared to the isotropic data in the 0-5\% centrality range. It shows that the power spectrum for AMPT have detailed structure due to flow but it's not the case for isotropic data.  

\section{Toy model jets}\label{sec:c}

\begin{figure}[ht]
\begin{minipage}[b]{0.45\linewidth}
            \centering
            \includegraphics[width=\textwidth]{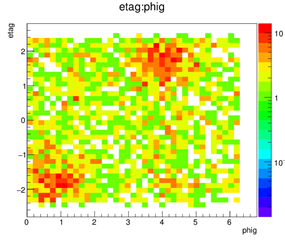}
            \caption{Distribution of the two added jets that depicts the high clustering of particles in the area in toy model simulation}
            \label{fig:e}
        \end{minipage}
        \hspace{0.5cm}
        \begin{minipage}[b]{0.45\linewidth}
            \centering
            \includegraphics[width=\textwidth]{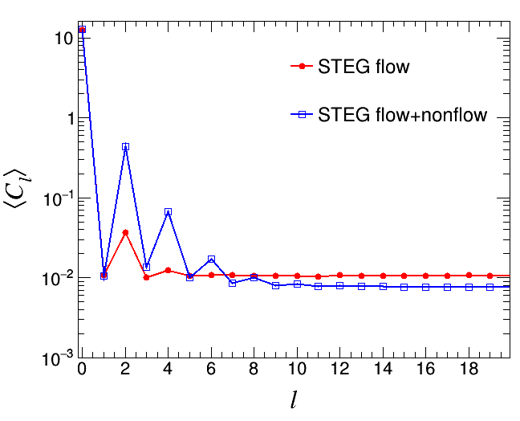}
            \caption{Figure depicting the larger contribution of the larger modes of $l$ compared to the events that do not contain the jets}
            \label{fig:f}
        \end{minipage}
\end{figure}

As the angular power spectrum can be used to represent flow, it may also be affected by the inclusion of jets. Jets used to probe QGP often appear in flow correlations as it is difficult to remove this nonflow from QGP expansion during analysis. However, simulations are able to provide some different ways to simulate the inclusion or exclusion of jets. In this case, toy models with particles produced similarity to the CMS detector acceptance are used to create two large jets in order to test the validity of the angular power spectrum. The jets were placed randomly as seen in Fig.~\ref{fig:e}. Because there is a clear clustering of the data causing the distribution to be much less uniform than the distribution without jet inclusion. It is clear that there will be a higher inclusion of higher $l$ modes. This is shown in Fig.~\ref{fig:f}. Though this was an exaggerated example of nonflow, it is clear that the angular power spectrum is affected by jets through the higher value of $l>1$ modes. 

\section{String melting analysis}\label{sec:d}

\begin{figure}[ht]
\begin{center}
\includegraphics[width=0.6\linewidth]{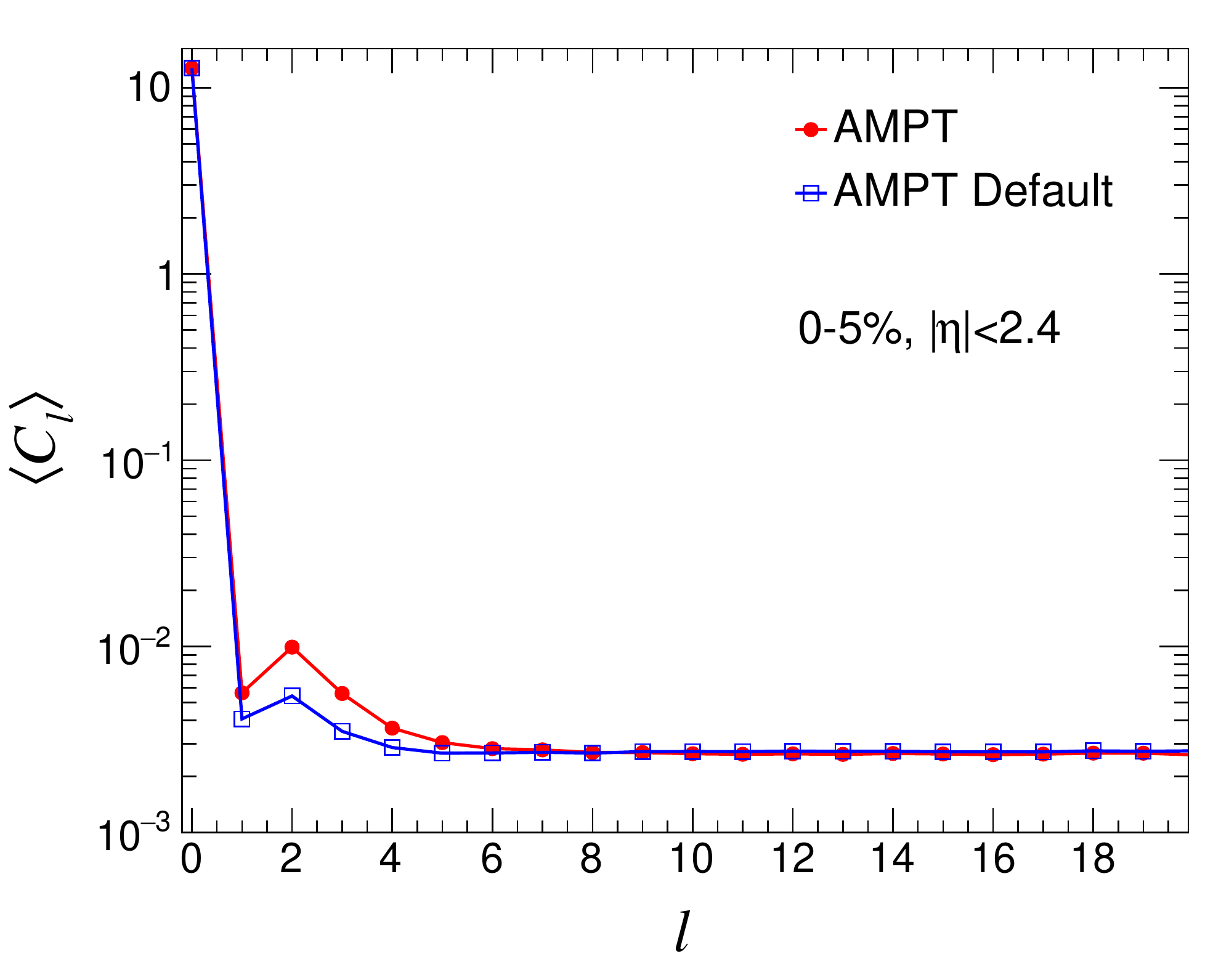}
\end{center}
\caption{\label{fig:subeventcartoon3} The smaller contribution of the $l=2$ mode is due to the lack of string melting which removes QGP indicators from the events}
\end{figure}

In the AMPT model, the initial parameter of string melting is included in order to facilitate interactions between partons after the collision. This is used to depict the expansion of QGP and subsequent hadronization accurately. However, when string melting is turned off, the hadronization still occurs, but there is partonic interactions. Therefore, it is also important to determine the effects this has on the angular power spectrum. Since studies on the angular power spectrum have thus far only included Pb-Pb collisions, which have been proven to contain indicators of QGP, it is imperative to understand the possible effects on the angular power spectrum when no QGP indicators are present. As seen in Fig.~\ref{fig:subeventcartoon3}, the $l=2$ mode diminishes when no QGP is present. This is the only even $l$ mode that depicts a peak when describing the AMPT data, therefore other even $l$ modes may become diminished if they are necessary to describe the distribution. However, it is possible that only the $l=2$ mode becomes smaller since this is the only mode that is used in the calculation of v2.

\section{Exclusion of m=0 mode}\label{sec:e}

\begin{figure}[ht]
\begin{center}
\includegraphics[width=0.6\linewidth]{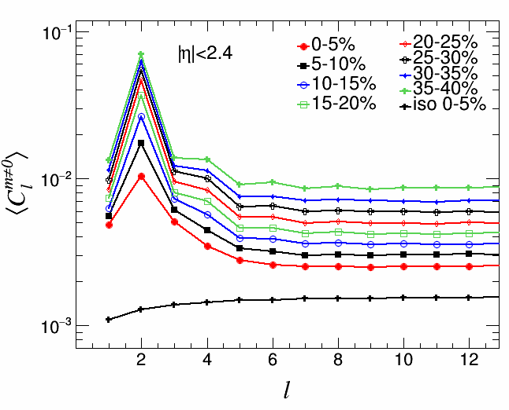}
\end{center}
\caption{\label{fig:subeventcartoon1} The contribution of m=0 terms gets rid of most of the higher $l$ mode dependence and removes $l=0$ mode entirely. However, there is still some fluctuation in the higher $l$ modes in less central events}
\end{figure}

Using the method proposed in Ref.~\cite{Machado:2018xvi}, in order to correct for edge effect introduced from detector acceptance (psuedorapidity range) cuts off, a new definition of the angular power system without the inclusion of $m=0$ modes will be tested. This modified angular power spectrum is defined as: 
\ball{aa6}
C_l^{m\neq0} = \frac{1}{2l+1}\sum_{m=-l}^{m=l}|a_{lm}|^2-\frac{|a_{l0}|^2}{2l+1}
\gal
As seen in Fig.~\ref{fig:subeventcartoon1}, because of this correction, the $l=2$ mode appears higher than the larger $l$ modes as compared to the initial power spectrum, and is much more evident in middle centrality ranges. 

\section{Flow Correlations}\label{sec:f}

Flow harmonics $v_n$ is related to the particle $\phi$ distribution through the Fourier decomposition~\cite{Voloshin:1994mz,Poskanzer:1998yz}:
\ball{aa4}
\frac{dN}{d\phi} \propto \frac{1}{2\pi}[1+2\sum_{n=1}^{\infty}v_ncos(n(\phi-\psi_n))]\,
\gal
where $\psi_n$ is the symmetric plane for different n modes

By introducing the notations $a_{lm}$ and $b_{lm}$ with 
\ball{a5}
a_{l0} = b_{l0}\ \textrm{for} \ m=0 
\gal
\ball{aa5}
a_{lm} = b_{lm}v_{|m|}e^{im\psi_{|m|}} \ \textrm{for} \ m \neq 0
\gal
\ball{a7}
b_{lm} = \sqrt{\frac{2l+1}{4\pi}\frac{(l-m)!}{(l+m)!}}\int_{\theta_i}^{\theta_f}sin(\theta)P_{lm}(cos\theta)d\theta
\gal
the flow coefficients $v_n$ can be expressed as~\cite{Machado:2018xvi}:
\ball{aa7}
|v_n|^2 = \frac{(2n+1)}{2}\frac{C_n^{m\neq0}}{|b_{nn}|^2}\frac{|b_{00}|^2}{4\pi}
\gal

\begin{figure}[ht]
\begin{minipage}[b]{0.45\linewidth}
            \centering
            \includegraphics[width=\textwidth]{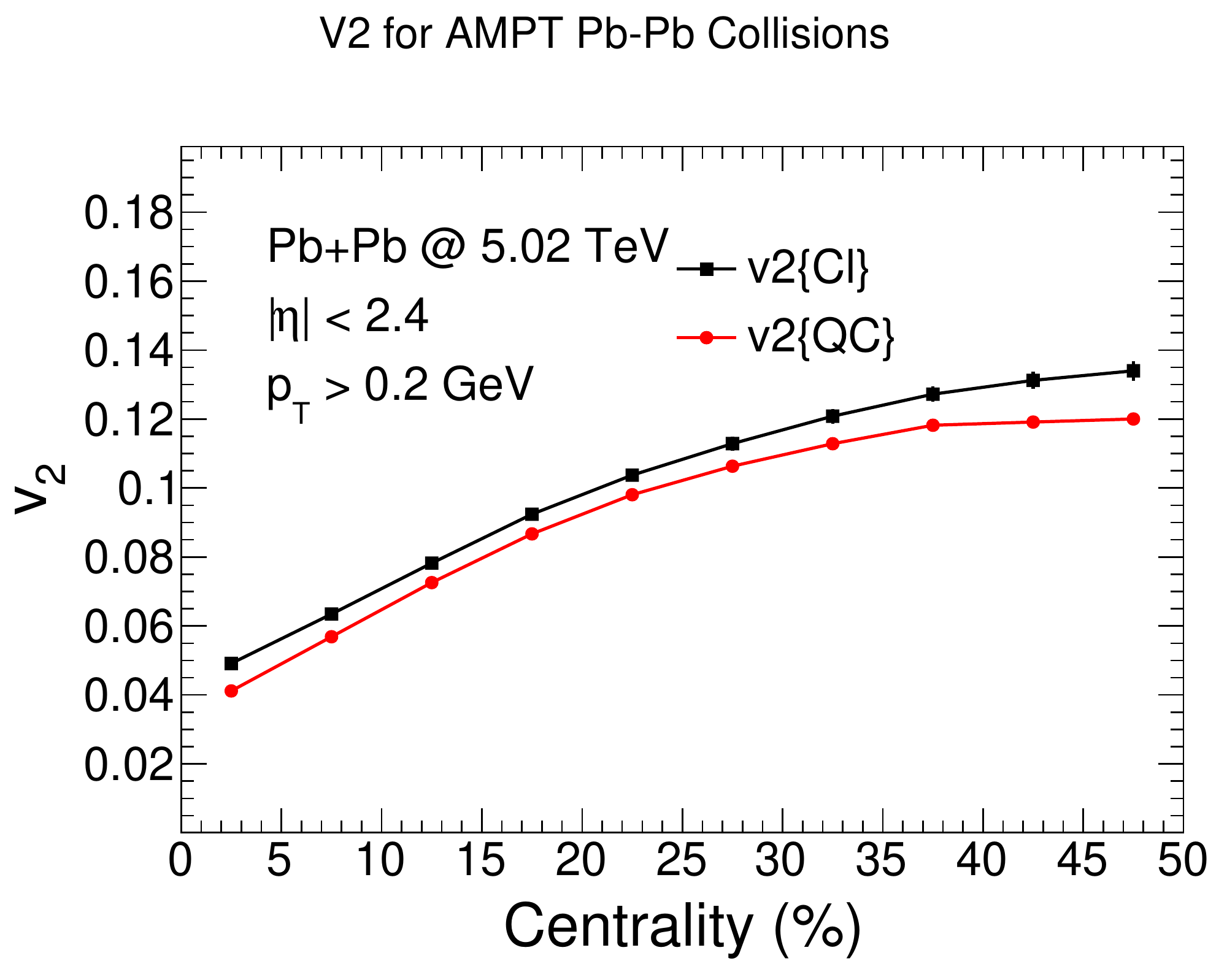}
            \caption{$v_2$ comparison for AMPT collisions for $p_T > 0.2$ GeV}
            \label{fig:g}
        \end{minipage}
        \hspace{0.5cm}
        \begin{minipage}[b]{0.45\linewidth}
            \centering
            \includegraphics[width=\textwidth]{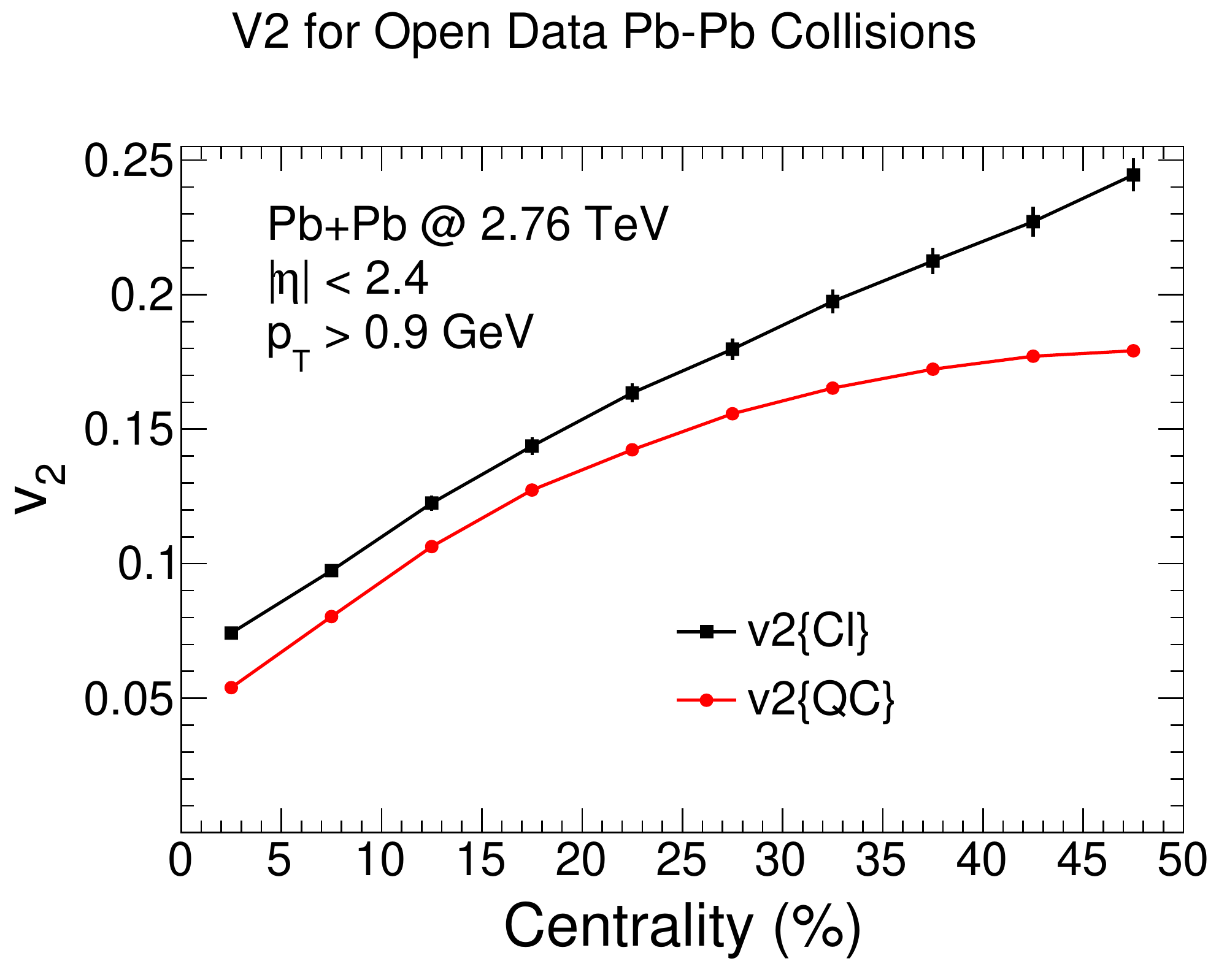}
            \caption{$v_2$ comparison for CMS open collision data with $p_T > 0.9$ GeV}
            \label{fig:h}
        \end{minipage}
\end{figure}

Using the equations of the $C_l^{m\neq0}$ from the power spectrum, the flow can be calculated through the equations above. This can then be compared with the Q-cumulants method of flow analysis for two particle correlations~\cite{Bilandzic:2010jr}. These coefficients have been labeled $v_2\{Cl\}$ and $v_2\{QC\}$ respectively as seen in Fig.~\ref{fig:g} and Fig.~\ref{fig:h}. Though both methods use two-particle correlations, the Q-cumulants only correlate the azimuthal angle while the angular power spectrum invloves both the azimuthal and polar angles. The $v_2$ calculations from both the angular power spectrum method and the Q-cumulants method agree well with a small margin of error and clearly follow the same trend for the AMPT data with $p_T > 0.2$. Though the open data does not converge as well as the AMPT data, it is still clear the results from both methods increase from central to middle central collisions.

\section{Discussion}\label{sec:s}

Overall, we have shown that analysis methods used originally for the Cosmic Microwave Background can be used with CMS open data. Detector limitations have been mostly eliminated, and the overlap between the two elliptic flow analysis methods for the AMPT data with $p_T>0.2$ GeV shows that this is a reliable method. In addition, we were able to look at the effects of nonflow through both toy model simulations and default AMPT setup without string melting. These models were able to depict the fluctuation in the angular power spectrum that can be attributed to nonflow. Because this study demonstrates that polar angle theta and azimuthal angle phi distributions can directly relate to elliptic flow, this brings up a question regarding the usefulness of HEALPix's Mollweide maps. If the distribution can accurately calculate the flow through HEALPix, can HEALPix maps somehow be used in differentiating between events that contain elliptic flow? This is a study that may be conducted through the use of machine learning algorithms and will be looked into in the future.

\acknowledgments

This work was supported in part by the U.S.\ Department of Energy under Grant No.\ DE-FG05-92ER40712.


\bibliography{apsvn_NPA}{}

\begin{thebibliography}{21}%
\makeatletter
\providecommand \@ifxundefined [1]{%
 \@ifx{#1\undefined}
}%
\providecommand \@ifnum [1]{%
 \ifnum #1\expandafter \@firstoftwo
 \else \expandafter \@secondoftwo
 \fi
}%
\providecommand \@ifx [1]{%
 \ifx #1\expandafter \@firstoftwo
 \else \expandafter \@secondoftwo
 \fi
}%
\providecommand \natexlab [1]{#1}%
\providecommand \enquote  [1]{``#1''}%
\providecommand \bibnamefont  [1]{#1}%
\providecommand \bibfnamefont [1]{#1}%
\providecommand \citenamefont [1]{#1}%
\providecommand \href@noop [0]{\@secondoftwo}%
\providecommand \href [0]{\begingroup \@sanitize@url \@href}%
\providecommand \@href[1]{\@@startlink{#1}\@@href}%
\providecommand \@@href[1]{\endgroup#1\@@endlink}%
\providecommand \@sanitize@url [0]{\catcode `\\12\catcode `\$12\catcode
  `\&12\catcode `\#12\catcode `\^12\catcode `\_12\catcode `\%12\relax}%
\providecommand \@@startlink[1]{}%
\providecommand \@@endlink[0]{}%
\providecommand \url  [0]{\begingroup\@sanitize@url \@url }%
\providecommand \@url [1]{\endgroup\@href {#1}{\urlprefix }}%
\providecommand \urlprefix  [0]{URL }%
\providecommand \Eprint [0]{\href }%
\providecommand \doibase [0]{http://dx.doi.org/}%
\providecommand \selectlanguage [0]{\@gobble}%
\providecommand \bibinfo  [0]{\@secondoftwo}%
\providecommand \bibfield  [0]{\@secondoftwo}%
\providecommand \translation [1]{[#1]}%
\providecommand \BibitemOpen [0]{}%
\providecommand \bibitemStop [0]{}%
\providecommand \bibitemNoStop [0]{.\EOS\space}%
\providecommand \EOS [0]{\spacefactor3000\relax}%
\providecommand \BibitemShut  [1]{\csname bibitem#1\endcsname}%
\let\auto@bib@innerbib\@empty
\bibitem [{\citenamefont {Adams}\ \emph {et~al.}(2005)\citenamefont {Adams}
  \emph {et~al.}}]{STAR:2005gfr}%
  \BibitemOpen
  \bibfield  {author} {\bibinfo {author} {\bibfnamefont {J.}~\bibnamefont
  {Adams}} \emph {et~al.} (\bibinfo {collaboration} {STAR}),\ }\href {\doibase
  10.1016/j.nuclphysa.2005.03.085} {\bibfield  {journal} {\bibinfo  {journal}
  {Nucl. Phys. A}\ }\textbf {\bibinfo {volume} {757}},\ \bibinfo {pages} {102}
  (\bibinfo {year} {2005})},\ \Eprint {http://arxiv.org/abs/nucl-ex/0501009}
  {arXiv:nucl-ex/0501009} \BibitemShut {NoStop}%
\bibitem [{\citenamefont {Adcox}\ \emph {et~al.}(2005)\citenamefont {Adcox}
  \emph {et~al.}}]{PHENIX:2004vcz}%
  \BibitemOpen
  \bibfield  {author} {\bibinfo {author} {\bibfnamefont {K.}~\bibnamefont
  {Adcox}} \emph {et~al.} (\bibinfo {collaboration} {PHENIX}),\ }\href
  {\doibase 10.1016/j.nuclphysa.2005.03.086} {\bibfield  {journal} {\bibinfo
  {journal} {Nucl. Phys. A}\ }\textbf {\bibinfo {volume} {757}},\ \bibinfo
  {pages} {184} (\bibinfo {year} {2005})},\ \Eprint
  {http://arxiv.org/abs/nucl-ex/0410003} {arXiv:nucl-ex/0410003} \BibitemShut
  {NoStop}%
\bibitem [{\citenamefont {Alver}\ \emph {et~al.}(2010)\citenamefont {Alver}
  \emph {et~al.}}]{Alver:2008aa}%
  \BibitemOpen
  \bibfield  {author} {\bibinfo {author} {\bibfnamefont {B.}~\bibnamefont
  {Alver}} \emph {et~al.} (\bibinfo {collaboration} {PHOBOS}),\ }\href
  {\doibase 10.1103/PhysRevC.81.024904} {\bibfield  {journal} {\bibinfo
  {journal} {Phys. Rev. C}\ }\textbf {\bibinfo {volume} {81}},\ \bibinfo
  {pages} {024904} (\bibinfo {year} {2010})},\ \Eprint
  {http://arxiv.org/abs/0812.1172} {arXiv:0812.1172 [nucl-ex]} \BibitemShut
  {NoStop}%
\bibitem [{\citenamefont {Arsene}\ \emph {et~al.}(2005)\citenamefont {Arsene}
  \emph {et~al.}}]{BRAHMS:2004adc}%
  \BibitemOpen
  \bibfield  {author} {\bibinfo {author} {\bibfnamefont {I.}~\bibnamefont
  {Arsene}} \emph {et~al.} (\bibinfo {collaboration} {BRAHMS}),\ }\href
  {\doibase 10.1016/j.nuclphysa.2005.02.130} {\bibfield  {journal} {\bibinfo
  {journal} {Nucl. Phys. A}\ }\textbf {\bibinfo {volume} {757}},\ \bibinfo
  {pages} {1} (\bibinfo {year} {2005})},\ \Eprint
  {http://arxiv.org/abs/nucl-ex/0410020} {arXiv:nucl-ex/0410020} \BibitemShut
  {NoStop}%
\bibitem [{\citenamefont {{ALICE Collaboration}}(2010)}]{Aamodt:2010pa}%
  \BibitemOpen
  \bibfield  {author} {\bibinfo {author} {\bibnamefont {{ALICE
  Collaboration}}},\ }\href {\doibase 10.1103/PhysRevLett.105.252302}
  {\bibfield  {journal} {\bibinfo  {journal} {Phys. Rev. Lett.}\ }\textbf
  {\bibinfo {volume} {105}},\ \bibinfo {pages} {252302} (\bibinfo {year}
  {2010})},\ \Eprint {http://arxiv.org/abs/1011.3914} {arXiv:1011.3914
  [nucl-ex]} \BibitemShut {NoStop}%
\bibitem [{\citenamefont {{ATLAS Collaboration}}(2012)}]{ATLAS:2012at}%
  \BibitemOpen
  \bibfield  {author} {\bibinfo {author} {\bibnamefont {{ATLAS
  Collaboration}}},\ }\href {\doibase 10.1103/PhysRevC.86.014907} {\bibfield
  {journal} {\bibinfo  {journal} {Phys. Rev. C}\ }\textbf {\bibinfo {volume}
  {86}},\ \bibinfo {pages} {014907} (\bibinfo {year} {2012})},\ \Eprint
  {http://arxiv.org/abs/1203.3087} {arXiv:1203.3087 [hep-ex]} \BibitemShut
  {NoStop}%
\bibitem [{\citenamefont {{CMS Collaboration}}(2013)}]{Chatrchyan:2012ta}%
  \BibitemOpen
  \bibfield  {author} {\bibinfo {author} {\bibnamefont {{CMS Collaboration}}},\
  }\href {\doibase 10.1103/PhysRevC.87.014902} {\bibfield  {journal} {\bibinfo
  {journal} {Phys. Rev. C}\ }\textbf {\bibinfo {volume} {87}},\ \bibinfo
  {pages} {014902} (\bibinfo {year} {2013})},\ \Eprint
  {http://arxiv.org/abs/1204.1409} {arXiv:1204.1409 [nucl-ex]} \BibitemShut
  {NoStop}%
\bibitem [{\citenamefont {Luzum}\ and\ \citenamefont
  {Romatschke}(2009)}]{Luzum:2009sb}%
  \BibitemOpen
  \bibfield  {author} {\bibinfo {author} {\bibfnamefont {M.}~\bibnamefont
  {Luzum}}\ and\ \bibinfo {author} {\bibfnamefont {P.}~\bibnamefont
  {Romatschke}},\ }\href {\doibase 10.1103/PhysRevLett.103.262302} {\bibfield
  {journal} {\bibinfo  {journal} {Phys. Rev. Lett.}\ }\textbf {\bibinfo
  {volume} {103}},\ \bibinfo {pages} {262302} (\bibinfo {year} {2009})},\
  \Eprint {http://arxiv.org/abs/0901.4588} {arXiv:0901.4588 [nucl-th]}
  \BibitemShut {NoStop}%
\bibitem [{\citenamefont {Schenke}\ \emph {et~al.}(2011)\citenamefont
  {Schenke}, \citenamefont {Jeon},\ and\ \citenamefont
  {Gale}}]{Schenke:2010rr}%
  \BibitemOpen
  \bibfield  {author} {\bibinfo {author} {\bibfnamefont {B.}~\bibnamefont
  {Schenke}}, \bibinfo {author} {\bibfnamefont {S.}~\bibnamefont {Jeon}}, \
  and\ \bibinfo {author} {\bibfnamefont {C.}~\bibnamefont {Gale}},\ }\href
  {\doibase 10.1103/PhysRevLett.106.042301} {\bibfield  {journal} {\bibinfo
  {journal} {Phys. Rev. Lett.}\ }\textbf {\bibinfo {volume} {106}},\ \bibinfo
  {pages} {042301} (\bibinfo {year} {2011})},\ \Eprint
  {http://arxiv.org/abs/1009.3244} {arXiv:1009.3244 [hep-ph]} \BibitemShut
  {NoStop}%
\bibitem [{\citenamefont {Heinz}\ and\ \citenamefont
  {Snellings}(2013)}]{Heinz:2013th}%
  \BibitemOpen
  \bibfield  {author} {\bibinfo {author} {\bibfnamefont {U.}~\bibnamefont
  {Heinz}}\ and\ \bibinfo {author} {\bibfnamefont {R.}~\bibnamefont
  {Snellings}},\ }\href {\doibase 10.1146/annurev-nucl-102212-170540}
  {\bibfield  {journal} {\bibinfo  {journal} {Ann. Rev. Nucl. Part. Sci.}\
  }\textbf {\bibinfo {volume} {63}},\ \bibinfo {pages} {123} (\bibinfo {year}
  {2013})},\ \Eprint {http://arxiv.org/abs/1301.2826} {arXiv:1301.2826
  [nucl-th]} \BibitemShut {NoStop}%
\bibitem [{\citenamefont {{Penzias}}\ and\ \citenamefont
  {{Wilson}}(1965)}]{1965ApJ142419P}%
  \BibitemOpen
  \bibfield  {author} {\bibinfo {author} {\bibfnamefont {A.~A.}\ \bibnamefont
  {{Penzias}}}\ and\ \bibinfo {author} {\bibfnamefont {R.~W.}\ \bibnamefont
  {{Wilson}}},\ }\href {\doibase 10.1086/148307} {\bibfield  {journal}
  {\bibinfo  {journal} {\apj}\ }\textbf {\bibinfo {volume} {142}},\ \bibinfo
  {pages} {419} (\bibinfo {year} {1965})}\BibitemShut {NoStop}%
\bibitem [{\citenamefont {Naselsky}\ \emph {et~al.}(2012)\citenamefont
  {Naselsky} \emph {et~al.}}]{Naselsky:2012nw}%
  \BibitemOpen
  \bibfield  {author} {\bibinfo {author} {\bibfnamefont {P.}~\bibnamefont
  {Naselsky}} \emph {et~al.},\ }\href {\doibase 10.1103/PhysRevC.86.024916}
  {\bibfield  {journal} {\bibinfo  {journal} {Phys. Rev. C}\ }\textbf {\bibinfo
  {volume} {86}},\ \bibinfo {pages} {024916} (\bibinfo {year} {2012})},\
  \Eprint {http://arxiv.org/abs/1204.0387} {arXiv:1204.0387 [hep-ph]}
  \BibitemShut {NoStop}%
\bibitem [{\citenamefont {{CMS Collaboration}}(2020)}]{CMS2010HIAllPhysics}%
  \BibitemOpen
  \bibfield  {author} {\bibinfo {author} {\bibnamefont {{CMS Collaboration}}},\
  }\href@noop {} {\emph {\bibinfo {title} {{HIAllPhysics primary dataset in
  RECO format from the 2.76 TeV Pb-Pb run of 2010
  (/HIAllPhysics/HIRun2010-ZS-v2/RECO), DOI:10.7483/OPENDATA.305G.POEC}}}},\
  \bibinfo {type} {Open Data}\ (\bibinfo  {institution} {CERN Open Data
  Portal},\ \bibinfo {year} {2020})\BibitemShut {NoStop}%
\bibitem [{\citenamefont {Lin}\ \emph {et~al.}(2005)\citenamefont {Lin},
  \citenamefont {Ko}, \citenamefont {Li}, \citenamefont {Zhang},\ and\
  \citenamefont {Pal}}]{Lin:2004en}%
  \BibitemOpen
  \bibfield  {author} {\bibinfo {author} {\bibfnamefont {Z.-W.}\ \bibnamefont
  {Lin}}, \bibinfo {author} {\bibfnamefont {C.~M.}\ \bibnamefont {Ko}},
  \bibinfo {author} {\bibfnamefont {B.-A.}\ \bibnamefont {Li}}, \bibinfo
  {author} {\bibfnamefont {B.}~\bibnamefont {Zhang}}, \ and\ \bibinfo {author}
  {\bibfnamefont {S.}~\bibnamefont {Pal}},\ }\href {\doibase
  10.1103/PhysRevC.72.064901} {\bibfield  {journal} {\bibinfo  {journal} {Phys.
  Rev. C}\ }\textbf {\bibinfo {volume} {72}},\ \bibinfo {pages} {064901}
  (\bibinfo {year} {2005})},\ \Eprint {http://arxiv.org/abs/nucl-th/0411110}
  {arXiv:nucl-th/0411110} \BibitemShut {NoStop}%
\bibitem [{\citenamefont {Sarwar}\ \emph {et~al.}(2018)\citenamefont {Sarwar},
  \citenamefont {Singh},\ and\ \citenamefont {Alam}}]{Sarwar:2017iax}%
  \BibitemOpen
  \bibfield  {author} {\bibinfo {author} {\bibfnamefont {G.}~\bibnamefont
  {Sarwar}}, \bibinfo {author} {\bibfnamefont {S.~K.}\ \bibnamefont {Singh}}, \
  and\ \bibinfo {author} {\bibfnamefont {J.-e.}\ \bibnamefont {Alam}},\ }\href
  {\doibase 10.1142/S0217751X1850121X} {\bibfield  {journal} {\bibinfo
  {journal} {Int. J. Mod. Phys. A}\ }\textbf {\bibinfo {volume} {33}},\
  \bibinfo {pages} {1850121} (\bibinfo {year} {2018})},\ \Eprint
  {http://arxiv.org/abs/1711.03743} {arXiv:1711.03743 [nucl-th]} \BibitemShut
  {NoStop}%
\bibitem [{\citenamefont {Llanes-Estrada}\ and\ \citenamefont {Mu\~noz
  Martinez}(2018)}]{Llanes-Estrada:2016pso}%
  \BibitemOpen
  \bibfield  {author} {\bibinfo {author} {\bibfnamefont {F.~J.}\ \bibnamefont
  {Llanes-Estrada}}\ and\ \bibinfo {author} {\bibfnamefont {J.~L.}\
  \bibnamefont {Mu\~noz Martinez}},\ }\href {\doibase
  10.1016/j.nuclphysa.2017.11.005} {\bibfield  {journal} {\bibinfo  {journal}
  {Nucl. Phys. A}\ }\textbf {\bibinfo {volume} {970}},\ \bibinfo {pages} {107}
  (\bibinfo {year} {2018})},\ \Eprint {http://arxiv.org/abs/1612.05036}
  {arXiv:1612.05036 [hep-ph]} \BibitemShut {NoStop}%
\bibitem [{\citenamefont {Machado}\ \emph {et~al.}(2019)\citenamefont
  {Machado}, \citenamefont {Damgaard}, \citenamefont {Gaardh\o{}je},\ and\
  \citenamefont {Bourjau}}]{Machado:2018xvi}%
  \BibitemOpen
  \bibfield  {author} {\bibinfo {author} {\bibfnamefont {M.}~\bibnamefont
  {Machado}}, \bibinfo {author} {\bibfnamefont {P.~H.}\ \bibnamefont
  {Damgaard}}, \bibinfo {author} {\bibfnamefont {J.~J.}\ \bibnamefont
  {Gaardh\o{}je}}, \ and\ \bibinfo {author} {\bibfnamefont {C.}~\bibnamefont
  {Bourjau}},\ }\href {\doibase 10.1103/PhysRevC.99.054910} {\bibfield
  {journal} {\bibinfo  {journal} {Phys. Rev. C}\ }\textbf {\bibinfo {volume}
  {99}},\ \bibinfo {pages} {054910} (\bibinfo {year} {2019})},\ \Eprint
  {http://arxiv.org/abs/1812.07449} {arXiv:1812.07449 [hep-ph]} \BibitemShut
  {NoStop}%
\bibitem [{\citenamefont {G\'orski}\ \emph {et~al.}(2005)\citenamefont
  {G\'orski}, \citenamefont {Hivon}, \citenamefont {Banday}, \citenamefont
  {Wandelt}, \citenamefont {Hansen}, \citenamefont {Reinecke},\ and\
  \citenamefont {Bartelman}}]{Gorski:2004by}%
  \BibitemOpen
  \bibfield  {author} {\bibinfo {author} {\bibfnamefont {K.~M.}\ \bibnamefont
  {G\'orski}}, \bibinfo {author} {\bibfnamefont {E.}~\bibnamefont {Hivon}},
  \bibinfo {author} {\bibfnamefont {A.~J.}\ \bibnamefont {Banday}}, \bibinfo
  {author} {\bibfnamefont {B.~D.}\ \bibnamefont {Wandelt}}, \bibinfo {author}
  {\bibfnamefont {F.~K.}\ \bibnamefont {Hansen}}, \bibinfo {author}
  {\bibfnamefont {M.}~\bibnamefont {Reinecke}}, \ and\ \bibinfo {author}
  {\bibfnamefont {M.}~\bibnamefont {Bartelman}},\ }\href {\doibase
  10.1086/427976} {\bibfield  {journal} {\bibinfo  {journal} {Astrophys. J.}\
  }\textbf {\bibinfo {volume} {622}},\ \bibinfo {pages} {759} (\bibinfo {year}
  {2005})},\ \Eprint {http://arxiv.org/abs/astro-ph/0409513}
  {arXiv:astro-ph/0409513} \BibitemShut {NoStop}%
\bibitem [{\citenamefont {Voloshin}\ and\ \citenamefont
  {Zhang}(1996)}]{Voloshin:1994mz}%
  \BibitemOpen
  \bibfield  {author} {\bibinfo {author} {\bibfnamefont {S.}~\bibnamefont
  {Voloshin}}\ and\ \bibinfo {author} {\bibfnamefont {Y.}~\bibnamefont
  {Zhang}},\ }\href {\doibase 10.1007/s002880050141} {\bibfield  {journal}
  {\bibinfo  {journal} {Z. Phys. C}\ }\textbf {\bibinfo {volume} {70}},\
  \bibinfo {pages} {665} (\bibinfo {year} {1996})},\ \Eprint
  {http://arxiv.org/abs/hep-ph/9407282} {arXiv:hep-ph/9407282} \BibitemShut
  {NoStop}%
\bibitem [{\citenamefont {Poskanzer}\ and\ \citenamefont
  {Voloshin}(1998)}]{Poskanzer:1998yz}%
  \BibitemOpen
  \bibfield  {author} {\bibinfo {author} {\bibfnamefont {A.~M.}\ \bibnamefont
  {Poskanzer}}\ and\ \bibinfo {author} {\bibfnamefont {S.~A.}\ \bibnamefont
  {Voloshin}},\ }\href {\doibase 10.1103/PhysRevC.58.1671} {\bibfield
  {journal} {\bibinfo  {journal} {Phys. Rev. C}\ }\textbf {\bibinfo {volume}
  {58}},\ \bibinfo {pages} {1671} (\bibinfo {year} {1998})},\ \Eprint
  {http://arxiv.org/abs/nucl-ex/9805001} {arXiv:nucl-ex/9805001} \BibitemShut
  {NoStop}%
\bibitem [{\citenamefont {Bilandzic}\ \emph {et~al.}(2011)\citenamefont
  {Bilandzic}, \citenamefont {Snellings},\ and\ \citenamefont
  {Voloshin}}]{Bilandzic:2010jr}%
  \BibitemOpen
  \bibfield  {author} {\bibinfo {author} {\bibfnamefont {A.}~\bibnamefont
  {Bilandzic}}, \bibinfo {author} {\bibfnamefont {R.}~\bibnamefont
  {Snellings}}, \ and\ \bibinfo {author} {\bibfnamefont {S.}~\bibnamefont
  {Voloshin}},\ }\href {\doibase 10.1103/PhysRevC.83.044913} {\bibfield
  {journal} {\bibinfo  {journal} {Phys. Rev. C}\ }\textbf {\bibinfo {volume}
  {83}},\ \bibinfo {pages} {044913} (\bibinfo {year} {2011})},\ \Eprint
  {http://arxiv.org/abs/1010.0233} {arXiv:1010.0233 [nucl-ex]} \BibitemShut
  {NoStop}%
\end{thebibliography}%

\end{document}